\documentclass[fleqn,12pt]{article}
\usepackage{amsfonts,amssymb,epsfig,latexsym,epic}
\usepackage[nospace]{cite}

\renewcommand{\baselinestretch}{1.1}

\makeatletter
\newbox\slashbox \setbox\slashbox=\hbox{$/$}
\def\pFMslash#1{\setbox\@tempboxa=\hbox{$#1$}
  \@tempdima=0.5\wd\slashbox \advance\@tempdima 0.5\wd\@tempboxa
  \copy\slashbox \kern-\@tempdima \box\@tempboxa}

\makeatother

\newcommand{\Ga}{\alpha}
\newcommand{\Gb}{\beta}

\newcommand{\Gd}{\delta}
\newcommand{\Ge}{\epsilon}
\newcommand{\Geps}{\varepsilon}
\newcommand{\Gg}{\gamma}

\newcommand{\GT}{{\cal T}}

\newcommand{\GTh}{\Theta}

\newcommand{\cA}{{\scriptscriptstyle\cal A}}
\newcommand{\cB}{{\scriptscriptstyle\cal B}}

\newcommand{\cM}{{\scriptscriptstyle\cal M}}
\newcommand{\cN}{{\scriptscriptstyle\cal N}}
\newcommand{\cK}{{\scriptscriptstyle\cal K}}
\newcommand{\cL}{{\scriptscriptstyle\cal L}}
\newcommand{\cP}{{\scriptscriptstyle\cal P}}
\newcommand{\cQ}{{\scriptscriptstyle\cal Q}}

\newcommand{\CA}{{\cal A}}
\newcommand{\CB}{{\cal B}}
\newcommand{\CC}{{\cal C}}

\newcommand{\CJ}{{\cal J}}
\newcommand{\CM}{{\cal M}}
\newcommand{\CN}{{\cal N}}

\newcommand{\CL}{{\cal L}}

\newcommand{\CP}{{\cal P}}
\newcommand{\CQ}{{\cal Q}}
\newcommand{\CS}{{\cal S}}

\newcommand{\CV}{{\cal V}}

\newcommand{\Um}{{\underline{m}}}
\newcommand{\Un}{{\underline{n}}}
\newcommand{\Uk}{{\underline{k}}}
\newcommand{\Ul}{{\underline{l}}}

\newcommand{\TCB}{{\tilde{\cal{B}}}}
\newcommand{\TCC}{{\tilde{\cal{C}}}}

\newcommand{\TCL}{{\tilde{\cal{L}}}}
\newcommand{\TCP}{{\tilde{\cal{P}}}}
\newcommand{\TCQ}{{\tilde{\cal{Q}}}}
\newcommand{\TCS}{{\tilde{\cal{S}}}}
\newcommand{\TCV}{{\tilde{\cal{V}}}}

\newcommand{\ft}[2]{{\textstyle {\frac{#1}{#2}} }}
\newcommand{\dd}{\partial}

\newcommand{\be}{\begin{equation}}
\newcommand{\ee}{\end{equation}}
\newcommand{\ben}{\begin{displaymath}}
\newcommand{\een}{\end{displaymath}}
\newcommand{\ba}{\begin{eqnarray}}
\newcommand{\ea}{\end{eqnarray}}
\newcommand{\nn}{\nonumber}
\newcommand{\non}{\nonumber\\}
\newcommand{\bean}{\begin{eqnarray*}}
\newcommand{\eean}{\end{eqnarray*}}

\def\moth{\mathsurround=0pt}
\newdimen\zo \zo=0pt

\def\tick{\leaders\hrule height 0.5ex depth 0pt \hskip 0.5pt}
\def\upboxfill{$\moth \setbox\zo\hbox{\tick}%
  \hskip 2pt\hbox to 0pt{$\tick$\hss}\hrulefill \hbox to
6pt{$\tick$\hss}$}

\def\dtick{\leaders\hrule height .34pt depth .5ex \hskip 0.5pt}
\def\downboxfill{$\moth \setbox\zo\hbox{\dtick}%
  \hskip 2pt\hbox to 0pt{$\dtick$\hss}\hrulefill \hbox to
6pt{$\dtick$\hss}$}

\newcommand{\la}{\label}

\newcommand{\Ref}[1]{(\ref{#1})}

\makeatletter
\@addtoreset{equation}{section}
\makeatother

\newcommand{\EE}[2]{{{\rm E}_{{#1}({#2})}}}

\newcommand{\EG}[2]{{{\rm G}_{{#1}({#2})}}}

\newcommand{\SU}[1]{{{\rm SU}({#1})}}
\newcommand{\SO}[1]{{{\rm SO}({#1})}}

\newcommand{\cro}{\!\times\!}
\newcommand{\equ}{\!=\!}
\newcommand{\pls}{\!+\!}
\newcommand{\mis}{\!-\!}

\begin{document}

\thispagestyle{empty}

\begin{flushright}
AEI-2003-029 \\
ITP-UU-03/13 \\
SPIN-03/07
\end{flushright}

\bigskip

\begin{center}
{\bf\Large Chern-Simons vs.\ Yang-Mills gaugings}
\medskip

{\bf\Large in three dimensions}

\bigskip\bigskip

{\bf Hermann~Nicolai\footnote{\tt nicolai@aei.mpg.de} and
Henning~Samtleben\footnote{\tt h.samtleben@phys.uu.nl}}

\vspace{.3cm}
$^1${\small Max-Planck-Institut f{\"u}r Gravitationsphysik\\
  Albert-Einstein-Institut\\
  M\"uhlenberg 1, D-14476 Golm, Germany}

\vspace{.5cm}

$^2${\small Institute for Theoretical Physics \& Spinoza Institute\\
Utrecht University, Postbus 80.195 \\
3508 TD Utrecht, The Netherlands}

\end{center}
\renewcommand{\thefootnote}{\arabic{footnote}}
\setcounter{footnote}{0}
\bigskip
\medskip
\begin{abstract}
Recently, gauged supergravities in three dimensions with Yang-Mills
and Chern-Simons type interactions have been constructed. In this
article, we demonstrate that any gauging of Yang-Mills type with
semisimple gauge group ${\rm G}_0$, possibly including extra couplings
to massive Chern-Simons vectors, is equivalent on-shell to a pure
Chern-Simons type gauging with non-semisimple gauge group ${\rm
G}_0\ltimes {\rm T}\subset {\rm G}$, where ${\rm T}$ is a certain
translation group, and where ${\rm G}$ is the maximal global symmetry
group of the ungauged theory. We discuss several examples.
\end{abstract}

\vspace*{2.5cm}

\renewcommand{\thefootnote}{\arabic{footnote}}
\setcounter{footnote}{0}

\newpage

\section{Introduction}

As borne out by recent work, gauged supergravities in three spacetime
dimensions come in more guises than the corresponding models in
dimensions $D\geq 4$. This variety of theories is not least due to
the fact that in three dimensions vector and scalar fields are related
by duality (see e.g.~\cite{CJLP98} for a general discussion of such
dualities). The on-shell equivalence of scalars and vectors in three
dimensions is not only reflected in a much larger choice of gauge
groups but also in the (co-)existence of both Chern-Simons (CS) type
gaugings and Yang-Mills (YM) type gaugings, depending on which
fields carry the propagating bosonic degrees of freedom.

YM type gaugings can be obtained alternatively by direct construction,
by torus reduction of gauged supergravities in higher dimensions to
three dimensions, or by Kaluza Klein reduction on non-flat internal
manifolds to three space-time dimensions
(see~\cite{LuPoTo97,CvLuPo00,LuPoSe02} for examples of such
constructions).  Examples of CS type gaugings were first obtained for
$N\equ2$ supergravity with an abelian gauge
group~\cite{DKSS00}. Maximal ($N\equ16$) and half maximal ($N\equ8$)
gauged supergravities of CS type with various compact and non-compact
semisimple gauge groups were constructed in~\cite{NicSam00} and
\cite{NicSam01b}, respectively (the relevant gauge groups are
subgroups of $\EE88$ for $N\equ16$ and $\SO{8,n}$ for $N\equ8$).  The
CS type supergravities are based on those versions of the ungauged
theories, in which all propagating bosonic degrees of freedom reside
in the scalar fields, and which therefore exhibit the largest global
symmetry group ${\rm G}$; the scalar fields parametrizing the coset
space manifold ${\rm G}/{\rm H}$. By contrast, YM type gaugings are
deformations of a dualized version of the ungauged theory where some
of the propagating bosonic degrees of freedom are carried by abelian
vector fields. The actual construction of the CS type gauged
Lagrangians is greatly facilitated by exploiting the group structure;
a simple and universal group-theoretical consistency condition
determines all the admissible gauge groups for an arbitrary number of
local supersymmetries, as first shown for the $N\equ16$ and $N\equ8$
theories \cite{NicSam00,NicSam01b}, and subsequently for all other
theories with $N<16$ \cite{dWHeSa03}.  The direct construction of the
YM type theories, on the other hand, seems more involved, because the
global symmetry group ${\rm G}$ is broken to a smaller group ${\rm
G}'\subset {\rm G}$, and furthermore the remaining scalars cannot be
assigned to a single coset space any more in general. The relative
simplicity of the CS type formulation in comparison with the YM type
formulation is evident from inspection of the resulting on-shell
equivalent Lagrangians~\Ref{LCS} and~\Ref{LYM_exta} below.

In this paper we establish, independently of the number $N$ of
local supersymmetries, that YM gaugings in three dimensions are in
fact equivalent on-shell to CS type gaugings, in the following
sense. The equations of motion of any gauged supergravity of YM
type with (semisimple) gauge group ${\rm G}_0$ coincide with the
equations of motion obtained from a corresponding CS type gauged
supergravity with non-semisimple gauge group ${\rm G}_0\ltimes
{\rm T}_\nu$, where ${\rm T}_\nu$ is a group of $\nu=\dim {\rm
G}_0$ (abelian) translations transforming non-trivially
under~${\rm G}_0$. Our second main result is the extension of this
construction to include couplings of massive CS vector fields to
the YM type Lagrangian. On the CS side these correspond to
additional nilpotent directions in the gauge group with a
particular algebra structure, see~\Ref{algebra_ext} below.

\begin{figure}[bt]
  \begin{center}
    \leavevmode
\begin{picture}(0,0)%
\includegraphics{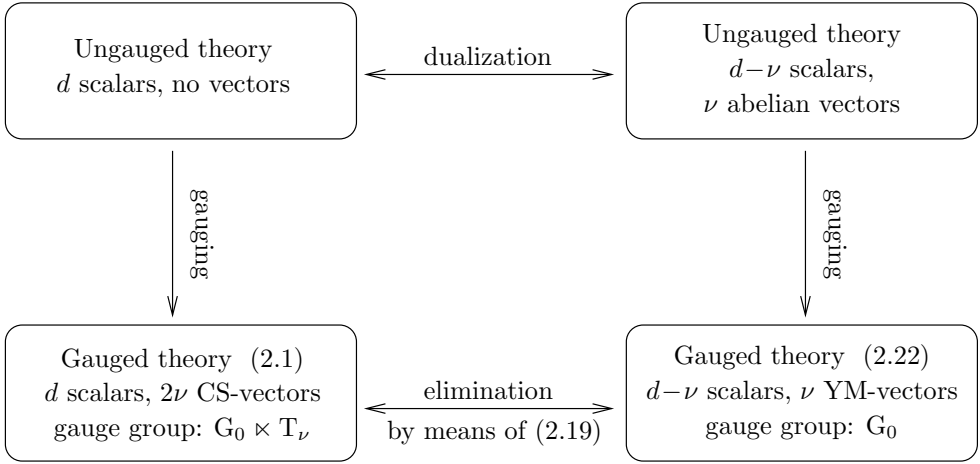}%
\end{picture}%
\setlength{\unitlength}{3947sp}%
\begingroup\makeatletter\ifx\SetFigFont\undefined%
\gdef\SetFigFont#1#2#3#4#5{%
  \reset@font\fontsize{#1}{#2pt}%
  \fontfamily{#3}\fontseries{#4}\fontshape{#5}%
  \selectfont}%
\fi\endgroup%
\begin{picture}(6129,2904)(1789,-4648)
\put(3001,-2911){\rotatebox{270.0}{\makebox(0,0)[lb]{\smash{\SetFigFont{10}{12.0}{\rmdefault}{\mddefault}{\updefault}gauging}}}}
\put(6976,-2911){\rotatebox{270.0}{\makebox(0,0)[lb]{\smash{\SetFigFont{10}{12.0}{\rmdefault}{\mddefault}{\updefault}gauging}}}}
\put(1951,-4246){\makebox(0,0)[lb]{\smash{\SetFigFont{10}{12.0}{\rmdefault}{\mddefault}{\updefault}\begin{tabular}{c}Gauged theory\,\, \Ref{LCS}\\$d$ scalars, $2\nu$ CS-vectors\\gauge group: ${\rm G}_0\ltimes {\rm T}_\nu$\end{tabular}}}}
\put(5931,-2206){\makebox(0,0)[lb]{\smash{\SetFigFont{10}{12.0}{\rmdefault}{\mddefault}{\updefault}\begin{tabular}{c}\phantom{M}Ungauged theory\\\phantom{M}$d\!-\!\nu$ scalars,\\\phantom{M}$\nu$ abelian vectors\end{tabular}}}}
\put(5731,-4231){\makebox(0,0)[lb]{\smash{\SetFigFont{10}{12.0}{\rmdefault}{\mddefault}{\updefault}\begin{tabular}{c}Gauged theory\,\, \Ref{LYM}\\$d\!-\!\nu$ scalars, $\nu$ YM-vectors\\gauge group: ${\rm G}_0$\end{tabular}}}}
\put(4431,-2131){\makebox(0,0)[lb]{\smash{\SetFigFont{10}{12.0}{\rmdefault}{\mddefault}{\updefault}dualization}}}
\put(2026,-2191){\makebox(0,0)[lb]{\smash{\SetFigFont{10}{12.0}{\rmdefault}{\mddefault}{\updefault}\begin{tabular}{c}Ungauged theory\\$d$ scalars, no vectors\end{tabular}}}}
\put(4431,-4231){\makebox(0,0)[lb]{\smash{\SetFigFont{10}{12.0}{\rmdefault}{\mddefault}{\updefault}elimination}}}
\put(4201,-4486){\makebox(0,0)[lb]{\smash{\SetFigFont{10}{12.0}{\rmdefault}{\mddefault}{\updefault}by means of \Ref{DPG}}}}
\end{picture}
\end{center}
  \caption{CS and YM gauged supergravity in three dimensions}
  \label{diag}
\end{figure}

Generally, the scalar fields of the CS gauged theory (whose number we
denote by $d$) parametrize a coset space ${\rm G}/{\rm H}$ with ${\rm
H}$ the maximal compact subgroup of ${\rm G}$. Here ${\rm G}$ is the
maximal global symmetry of the ungauged theory that can be achieved by
dualizing all propagating bosonic degrees of freedom into scalar
fields. Obviously, the group we wish to gauge must then satisfy ${\rm
G}_0\ltimes {\rm T}_\nu\subset {\rm G}$. Given a gauge group of this
kind, we show that $\nu$ scalar and $\nu$ vector fields may be
eliminated together, whereupon the theory turns into a YM gauged
theory with $(d\mis\nu)$ scalars and $\nu$ propagating vector fields
gauging the group ${\rm G}_0$. It is important that the scalar
potential is independent of the scalar fields in question and is
therefore not affected by this elimination procedure. After the
elimination, the $(d\mis\nu)$ scalars of the YM gauged theory in
general can no longer be uniformly described as a coset space; only
part of them can be assigned to a (smaller) coset space ${\rm G}'/{\rm
H}'$ with ${\rm G}'\subset {\rm G}$, ${\rm H}'\subset {\rm H}$ (of
course, the YM gauge group ${\rm G}_0$ must be contained in ${\rm
G}'$). Allowing also for some bosonic degrees of freedom to be
realized as massive CS vectors (see section~3), the following general
matching condition for the bosonic degrees of freedom must evidently
be satisfied
\ba
\label{matching}
d &=& \dim {\rm G}/{\rm H} ~=~
\#({\rm scalars}) + \#({\rm YM~vectors}) + \#({\rm
massive~CS~vectors})
\;.
\ea
The procedure relating the two types of theories is schematically
represented in figure~\ref{diag}.

The necessity of flat directions in the CS gauge group, and hence
of non-semisimple gauge groups, for the transmutation of a CS type
theory into a YM type theory may be understood by noting that only
those scalar fields on which the scalar potential of the gauged
theory does not depend can be dualized away and replaced by YM
vector fields. Hence the associated translations along these
directions on the target space manifold must be among the local
symmetries of the CS type Lagrangian. There exist numerous results
on non-semisimple gaugings in $D\geq 4$
\cite{Hull84a,ACFG00,ADFL02,Hull02,dWSaTr02}, but non-semisimple
gauge groups have not been considered so far in the context of CS
gauged supergravities in three dimensions. These will be treated
in detail in a forthcoming publication \cite{FiNiSa03}, so we here
only note that there are two methods to search for them.  The
first is to directly solve the group-theoretical consistency
conditions of~\cite{NicSam00,NicSam01b,dWHeSa03}. For the second,
one performs an infinite ``boost'' on a known admissible
semisimple gauge group with a suitable non-compact element of the
global symmetry group ${\rm G}$; this boost must be accompanied by
a singular rescaling of the coupling constants, which is adjusted
in such a way that the embedding tensor $\Theta$ has a finite
limit.

In summary, the set of CS gauged supergravities in three
dimensions with non-semisimple gauge groups contains all the known
types of YM gauged supergravities. Because there exist numerous
admissible semisimple gauge groups, that cannot be related to YM
type gaugings, it follows that the CS gauged supergravities
encompass a much larger class of models than those of YM type.
Combining the classification of ungauged three-dimensional
theories~\cite{dWToNi93} with the group-theoretical results on the
existence of CS gaugings~\cite{NicSam00,NicSam01b,dWHeSa03} thus
provides a straightforward route to constructing gauged
supergravity theories in three dimensions for a given field
content, gauge group and number of supersymmetries.

The paper is organized as follows. In section~\ref{CSYM} we provide
details of the correspondence outlined in figure~\ref{diag},
exhibiting the YM type gaugings as a particular subclass of CS gauged
theories.  Section~\ref{MVF} describes the generalization to coupling
additional massive vector fields. We close by briefly discussing
several examples, including the compactifications of type I, IIA, IIB
supergravity in $D=10$ on AdS$_3\times S^7$ and the six-dimensional
supergravity on AdS$_3\times S^3$.

\section{Chern-Simons vs.\ Yang-Mills gauging}
\label{CSYM}

The bosonic Lagrangian of a CS gauged supergravity theory in three
dimensions is always of the form (see \cite{NicSam00,NicSam01b} for
our conventions and notations; we use the metric $(+--)$)
\ba
e^{-1}\CL &=& -\ft1{4} R
+ e^{-1}g\CL_{\rm CS}
+ \ft1{4} g^{\mu\nu}\,\CP^A_\mu\CP^A_\nu - W + \dots  \;.
\label{LCS}
\ea
The dots stand for fermionic terms which we will here ignore because
they are not relevant for the argument we are going to present; see
however \cite{NicSam00,NicSam01b,dWHeSa03} for further details
concerning the fermionic Lagrangian and the supersymmetry
variations. The first term in \Ref{LCS} is just the usual Einstein
term, while the second is the CS Lagrangian
\ba
\CL_{\rm CS}&=&
\ft14 \Geps^{\mu\nu\rho} B^\cM_\mu \,\GTh_{\cM\cN}\,
\left( \dd_\nu B^\cN_\rho
+ \ft13 g\,
f^{\cN\cP}{}_{\cL}\,\GTh_{\cP\cK}\,B^\cK_\nu B^\cL_\rho\right)
\;.
\ea
with the constant symmetric embedding tensor $\GTh_{\cM\cN}$, which
characterizes the CS gauge group, see~\Ref{J} below. Note that
$\CL_{\rm CS}$ comes with a factor $g$, the gauge coupling constant.
$g\rightarrow 0$ describes the (smooth) limit to the ungauged theory.

To explain the remaining two terms in \Ref{LCS} we recall that the $d$
scalar fields parametrize a coset space ${\rm G}/{\rm H}$ where ${\rm
H}$ is the maximal compact subgroup of ${\rm G}$. Of course, the
choice of possible coset spaces depends on the number $N$ of local
supersymmetries and becomes more and more restricted with increasing
$N$ \cite{dWToNi93}. Explicitly, the scalar fields are described by a
group valued matrix $\CS\in {\rm G}$ such that the current
\ba
\CQ_\mu + \CP_\mu &\equiv& \CS^{-1}
\left( \dd_\mu + g\,\GTh_{\cM\cN}\,B^\cM_\mu\,t^\cN \right) \CS
\;,
\la{J}
\ea
takes values in the associated Lie algebra $\mathfrak{g}$, with a
suitably normalized basis $\{ t^\cM \}$, where $\CM, \CN = 1, \dots,
{\rm dim}\,{\rm G}$. The quantity $\CP_\mu=\CP_\mu^A t^A$ appearing
in the kinetic term of the scalar fields in the Lagrangian~\Ref{LCS}
is the projection of this current onto the noncompact part of
$\mathfrak{g}$, which is spanned by the generators $\{ t^A \}$ with
labels $A, B, \dots$. The compact part $\CQ_\mu$, on the other hand,
serves as a (composite) connection for the maximal compact subgroup
${\rm H}$ and governs the scalar-fermion couplings of the ungauged
theory. The embedding tensor $\GTh_{\cM\cN}$ describes the coupling of
the vector fields to the generators of the action of the symmetry
group, hence the embedding of the CS gauge group into~${\rm G}$.

The potential $W$ is likewise a function of the scalar fields
$\CS$. More specifically, it is a quadratic polynomial in the entries
of the $T$-tensor $T_{\cA\cB}$ which in turn is given in terms of the
matrix $\CV^\cM{}_\cA$ representing the group element $\CS$ in the
adjoint representation:
\be
T_{\cA\cB} ~\equiv~ \GTh_{\cM\cN}\,\CV^\cM{}_\cA
\CV^\cN{}_\cB \;,\qquad
\CV^\cM{}_\cA\,t^\cA ~\equiv~ \CS^{-1} t^\cM \CS  \;.
\la{TV}
\ee
The exact dependence of $W$ on $T_{\cA\cB}$ as well as the possible
gauge groups and their embedding matrices $\GTh_{\cM\cN}$ can be
found in~\cite{NicSam00,NicSam01b,dWHeSa03} and is completely
determined by supersymmetry.

The above Lagrangian is invariant under the (infinitesimal) gauge
transformations
\ba
\delta \CS &=& - g\,\GTh_{\cM\cN}\,\Lambda^\cM\,t^\cN\,\CS \;,
\qquad\quad
\delta B^\cM_\mu ~=~  \dd_\mu \Lambda^\cM
+ g\,f^{\cM\cP}{}_\cL\,\GTh_{\cP\cK}\,B_\mu^\cK\,\Lambda^\cL
\;.
\la{gauge}
\ea

Variation of the Lagrangian \Ref{LCS} with respect to the vector
fields gives rise to the first order duality equations (again
omitting fermionic contributions):
\ba
\GTh_{\cM\cN}\, \CB^\cN_{\mu\nu} &\equiv&
\GTh_{\cM\cN}\,\left(
2\dd^{\vphantom{\cN}}_{[\mu} B^\cN_{\nu]} +
g\,\GTh_{\cK\cP}\, f^{\cN\cK}{}_\cQ \,
B^\cP_\mu B^\cQ_\nu \right) ~=~
e \Geps_{\mu\nu\rho} \,
\GTh_{\cM\cN}\,\CV^\cN{}_A \, \CP^{A\rho}
\;.
\la{duality}
\ea

To proceed we now assume a very particular type of gauge group,
namely a non-semisimple group of the form ${\rm G}_0 \ltimes {\rm
T}_\nu \subset {\rm G}$, where ${\rm T}_\nu$ is a set of $\nu=\dim
{\rm G}_0$ translations transforming in the adjoint representation
of ${\rm G}_0$. A systematic discussion and representative
examples of such gaugings will be given in \cite{FiNiSa03}.
Denoting the generators of $\mathfrak{g}_0 \equiv {\rm Lie}\, {\rm
G}_0$ by $\{ \CJ^m \equiv t^m \}$ and those of
$\mathfrak{t}_\nu\equiv {\rm Lie}\, {\rm T}_\nu$ by $\{ \GT^\Um
\equiv t^\Um \}$, respectively, where both $m$ and $\Um$ range
over $1,\dots , \nu \equiv {\rm dim }\, {\rm G}_0$, we have the
commutation relations
\ba
\left[ \CJ^m ,\CJ^n \right] &=& f^{mn}{}_k\,\CJ^k \;,\qquad
\left[ \CJ^m ,\GT^\Un \right] ~=~ f^{mn}{}_k\, \GT^\Uk \;,\qquad
\left[ \GT^\Um ,\GT^\Un \right] ~=~ 0 \;,\qquad
\la{algebra}
\ea
where $f^{mn}{}_k$ denote the structure constants of ${\rm G}_0$, so
the translation generators transform in the adjoint of ${\rm G}_0$.
It can now be shown that for this particular choice of gauge group,
a consistent gauging is possible only if the embedding tensor
$\GTh_{\cM\cN}$ is of the form
\ba
g\,\GTh_{m\Un} &=& g\,\GTh_{\Um n} ~=~ g_1\,\eta_{mn} \;,
\qquad
g\,\GTh_{\Um\Un} ~=~ g_2\,\eta_{mn} \;,
\la{Theta}
\ea
with all remaining components equal to zero. Here $\eta_{mn}$ is
the Cartan-Killing form on~$\mathfrak{g}_0$. The
ansatz~\Ref{Theta} is uniquely fixed by demanding invariance of
$\GTh_{\cM\cN}$ under the gauge group~\Ref{algebra}. In
particular, this invariance requires $\Theta_{mn}=0$; happily,
this is also the condition needed for our elimination procedure to
work. The real constants $g_1$, $g_2$ in general cannot be freely
chosen, but are determined as a function of the one free gauge
coupling constant $g$ by how ${\rm G}_0 \ltimes {\rm T}_\nu$ is
embedded in ${\rm G}$, and by the fact that the embedding
tensor~\Ref{Theta} must satisfy the group-theoretical identities
of~\cite{NicSam00,NicSam01b,dWHeSa03} to ensure compatibility with
supersymmetry. After the elimination procedure we are about to
describe, the coupling $g_1$ will play the role of the YM gauge
coupling constant while $g_2$ corresponds to an inequivalent
deformation of the theory by an additional CS term.

We denote the vector fields associated with ${\rm G}_0$ and ${\rm
T}_\nu$ by $C_\mu^m \equiv B_\mu^m$ and $A_\mu^m\equiv B_\mu^\Um$,
respectively. Their transformation properties follow from \Ref{gauge}:
\ba
\Gd A_\mu^m &=&
\dd_\mu \Lambda^{\Um} +g_1\,f^m{}_{kl}\,A^k_{\mu}\,\Lambda^\Ul
\;,
\non[1ex]
\Gd C_\mu^m &=& \dd_\mu \Lambda^m +
    g_1 f^m{}_{kl}\, C_\mu^k \Lambda^\Ul +
  f^m{}_{kl}\, A_\mu^k (g_1 \Lambda^l + g_2 \Lambda^\Ul)
\;,
\la{gtvec}
\ea
where $f^m{}_{kl} \equiv \eta_{nk} {f^{mn}}_l$. The associated field
strengths can be read off from~\Ref{duality}; they are
\ba
\CA^m_{\mu\nu} ~\equiv~ \CB^\Um_{\mu\nu} &=& \dd_\mu A^m_\nu -
\dd_\nu A^m_\mu + g_1\,f^m{}_{kl}\,A^k_\mu A^l_\nu \;,\non[1ex]
\CC^m_{\mu\nu} ~\equiv~ \CB^m_{\mu\nu} &=& \dd_\mu C^m_\nu
- \dd_\nu C^m_\mu + 2\,g_1\,f^m{}_{kl}\,C^k_{[\mu} A^l_{\nu]} +
g_2\,f^m{}_{kl}\,A^k_\mu A^l_\nu  \;.
\la{FG}
\ea
Next we split off the translation part from the scalar field matrix
$\CS$
\ba
\CS (\phi, \tilde\phi) &\equiv& e^{\phi_m\,\GT^{\Um}}\,
\TCS (\tilde\phi) \;,
\la{Stilde}
\ea
in terms of $\nu$ scalars $\phi_m$ associated with the translation
generators $\{ \GT^\Um \}$, and the remaining scalar fields
$\tilde\phi$ some of which coordinatize the YM coset manifold ${\rm
G}'/{\rm H}'$. What is important is that the matrix $\TCS$ no longer
depends on the translational degrees of freedom $\phi_m$. Defining the
modified field strength
\ba
\tilde\CC^m_{\mu\nu} &\equiv&
\CC^m_{\mu\nu} - f^{mn}{}_k\, \phi_n \CA^k_{\mu\nu} \;,
\la{modC}
\ea
we have the transformation properties
\ba
\delta \CA^m_{\mu\nu} &=&
g_1\,f^m{}_{kl}\, \CA^k_{\mu\nu}\,\Lambda^\Ul
\;,
\qquad \delta \TCS ~=~ - g_1 \eta_{mn} \Lambda^{\Um}\, \CJ^n \TCS \;,
\non
\delta \tilde\CC^m_{\mu\nu} &=& g_1\,f^m{}_{kl}\,
\tilde\CC^k_{\mu\nu}\,\Lambda^\Ul
\;,
\qquad \delta \phi_m ~=~ - g_1 \eta_{mn}  \Lambda^n -
\left( g_2 \eta_{mn}  + g_1 f^l{}_{mn}\phi_l \right) \Lambda^\Un
\;,
\la{gaugeexp}
\ea
from \Ref{gauge}. Note that the fields $\phi_m$ and $C^m_{\mu}$ are the
only ones to transform with the translation parameters $\Lambda^m$,
and that $\phi_m$ is shifted under such transformation and hence could
be gauged away altogether. Accordingly, we define the quantities
\ba
\TCV^\cM{}_\cA\, t^\cA &\equiv& \TCS^{-1} t^\cM \TCS \;, \non
\TCQ_\mu + \TCP_\mu  &\equiv& \TCS^{-1}
\left( \dd_\mu + g_1 \eta_{mn}\,A^m_\mu\,\CJ^n \right) \TCS
\;,
\la{VT}
\ea
which do not depend on the $\phi_m$ either. It is then easy to check
that, for $\CM = m, \Um$
\be
\TCV^\Um{}_\cA=\CV^\Um{}_\cA  \;\; , \qquad \TCV^m{}_\cA =
\CV^m{}_\cA - f^{mn}{}_k\, \phi_n\,\CV^\Uk{}_\cA
\;.
\ee
As expected, the $T$-tensor \Ref{TV} does not depend on $\phi_m$.
This follows from the fact that by construction it is gauge invariant,
and more specifically invariant under the (local) $\Lambda^m$
translations on $\phi_m$, see~\Ref{gaugeexp}, but it is also easy to
verify directly that
\ba
T_{\cA\cB} &=& \GTh_{\cM\cN}\,\CV^\cM{}_\cA \CV^\cN{}_\cB ~=~
\GTh_{\cM\cN}\,\TCV^\cM{}_\cA \TCV^\cN{}_\cB \;,
\ea
with $\GTh_{\cM\cN}$ from \Ref{Theta}. Consequently, the scalar
potential $W$ in \Ref{LCS} does not depend on~$\phi_m$. After a little
algebra, the current $\CQ_\mu + \CP_\mu$ can be rewritten as
\ba
\CQ_\mu + \CP_\mu &=& \TCQ_\mu + \TCP_\mu +
\left(\dd_\mu \phi_m +
\eta_{mn}\left(g_1 C_\mu^n + g_2 A_\mu^n \right) + g_1
f^k{}_{mn}\,A_\mu^n\,\phi_k \right) \TCV^\Um{}_\cA t^\cA
\non[1ex]
&\equiv&
\TCQ_\mu + \TCP_\mu  + D_\mu \phi_m \, \TCV^\Um{}_\cA t^\cA
\;.
\la{PQtild}
\ea
with the definition of the covariant derivative in accordance
with~\Ref{gaugeexp}. The first order duality equations \Ref{duality}
for the gauge group \Ref{algebra}, \Ref{Theta}, take the form
\ba
\CA^m_{\mu\nu} &=& e \Geps_{\mu\nu\rho} \,
\TCV^\Um{}_A \left(\TCP^{A\rho} + \TCV^\Un{}_A\,D^\rho\phi_n \right)
\;,\non[1ex]
\TCC^m_{\mu\nu} &=& e \Geps_{\mu\nu\rho} \,
\TCV^m{}_A \left(\TCP^{A\rho} + \TCV^\Un{}_A\,D^\rho\phi_n \right) \;,
\la{dualityFG}
\ea
with the modified field strength $\TCC^m_{\mu\nu}$
from~\Ref{modC}. Hence they may be formulated exclusively in terms of
objects that are invariant under $\Lambda^m$ and transform covariantly
under $\Lambda^\Um$.

Our aim is now to eliminate all $\phi_m$ dependence from the equations
of motion. To this end, assume the matrix
$M^{mn}\equiv\TCV^\Um{}_A\TCV^\Un{}_A$ to be invertible with inverse
$M_{mn}$.
This allows to solve equations \Ref{dualityFG} for $D_\mu \phi_m$ and
$\tilde\CC^m_{\mu\nu}$
\ba
e\Geps_{\mu\nu\rho} \,D^\rho \phi_m &=&
M_{mn} \,\CA^n_{\mu\nu} - e\Geps_{\mu\nu\rho}
\,M_{mn}\,\TCV^{\Un}{}_A\,\TCP^{A\rho}
\;,\non
\tilde\CC^m_{\mu\nu} &=& e\Geps_{\mu\nu\rho}
\left(\TCV^m{}_A-\TCV^m{}_B\TCV^\Uk{}_B\,M_{kl}\,\TCV^\Ul{}_A \right)
\TCP^{A\rho} + \TCV^m{}_A\TCV^\Uk{}_A\,M_{kn}\,\CA^n_{\mu\nu} \;.
\la{DPG}
\ea
These equations can now be used to eliminate both $\phi_m$ and
$C_\mu^m$ from the theory. Solubility of the first equation
in~\Ref{DPG} implies an integrability condition on the r.h.s.\ which
is straightforwardly computed using
\ba
\left[ D_\mu ,D_\nu \right] \phi_m &=&
\eta_{mn}
\left( g_1 \tilde\CC^n_{\mu\nu} + g_2 \CA^n_{\mu\nu} \right) \;,
\ea
and leads to the following second order field equation for the vector
fields $A_\mu^m$
\ba
D^\nu\left(M_{mn} \CA^{n}_{\mu\nu} \right) &=&
e \Geps_{\mu\nu\rho}\,D^\nu \left(M_{mn} \TCV^\Un{}_A \TCP^{A\rho}
\right)
+ g_1\,\eta_{mn} \TCV^n{}_A \left(
\delta_{AB} -  \TCV^\Uk{}_A M_{kl} \TCV^\Ul{}_B \right) \TCP^B_\mu
\non
&&{}+\ft12 e \Geps_{\mu\nu\rho} \left(
g_2\eta_{mn} + g_1 \eta_{mk} \TCV^k{}_A \TCV^\Ul{}_A M_{ln} \right)
\CA^{n \nu\rho}
\;.
\la{YM}
\ea
Hence this field equation is equivalent to the set of first order
equations \Ref{dualityFG} while the fields $\phi_m$ and $C^m_\mu$ are
completely decoupled and may be restored
from~\Ref{DPG}. (Integrability of the second equation in~\Ref{DPG} in
addition requires also part of the scalar field equations.)
Equation~\Ref{YM} may be derived from the Lagrangian
\ba
e^{-1}\TCL &=& -\ft1{4} R - e^{-1}g_2 \TCL_{\rm CS}(A)
-\ft18 \,M_{mn}\,\CA^{m\mu\nu}\CA^n_{\mu\nu}
+ \ft1{4} g^{\mu\nu}\,G_{AB}\,\TCP^A_\mu\TCP^B_\nu - W
\non[1ex]
&&{}
+\ft14 e^{-1} \,
\Geps^{\mu\nu\rho}\,M_{mn}\,\TCV^\Un{}_A\,\CA^{m}_{\mu\nu} \,
\TCP^A_\rho
\;,
\label{LYM}
\ea
with
\ba
G_{AB} &\equiv& \delta_{AB} -  \TCV^\Um{}_A M_{mn} \TCV^\Un{}_B
\;,\qquad M_{mn} ~\equiv~ (\TCV^\Um{}_A\TCV^\Un{}_A )^{-1}
\;,\non[1ex]
\TCL_{\rm CS}(A) &=&
\ft14\,\Geps^{\mu\nu\rho} A^m_\mu\, \eta_{mn}\,
\left( \dd_\nu A^n_\rho
+ \ft13 g_1 f^{n}{}_{kl}\,A^k_\nu A^l_\rho\right)
\;.
\la{LYMa}
\ea
It requires a little more work to show that the scalar field equations
derived from \Ref{LYM} reproduce those
descending from \Ref{LCS} upon eliminating $D_\mu\phi_m$ by means
of~\Ref{DPG}. Note that the metric $G_{AB}$ on the scalar target space
is degenerate along the $\nu$ directions $\TCV^\Um{}_A$. This just
means that the elimination of the scalar fields $\phi_m$ has
effectively reduced the dimension of the scalar manifold in
\Ref{LYM} by $\nu$. The Chern-Simons term in \Ref{LYM} collects
only part of the terms from the corresponding term in~\Ref{LCS}.  The
scalar potentials $W$ in \Ref{LCS} and \Ref{LYM} coincide.  The
resulting YM type theory thus has $d\mis\nu$~scalar fields and $\nu$
propagating Yang-Mills vectors. The residual gauge group ${\rm G}_0$
acts canonically as
\ba
\delta \TCS ~=~ - g_1 \eta_{mn} \Lambda^{\Um}\, \CJ^n \TCS \;, \qquad
\delta A^m_{\mu} &=&
\dd_\mu \Lambda^{\Um} +g_1\,f^m{}_{kl}\,A^k_{\mu}\,\Lambda^\Ul \;.
\ea
The fermionic part of the Lagrangian \Ref{LYM} as well as the
supersymmetry transformation rules may be directly obtained from those
of~\Ref{LCS} upon eliminating $D_\mu\phi_m$ and $\CC^m_{\mu\nu}$ by
means of \Ref{DPG}.

In the (smooth) limit $g_1\rightarrow0$, $g_2\rightarrow0$, the
Lagrangian \Ref{LYM} reduces to the ungauged theory with $d-\nu$
scalar fields and $\nu$ abelian vectors. The metrics $M_{mn}$ and
$G_{AB}$ in the kinetic terms remain unchanged in this limit. As
anticipated above, the two constants $g_1$ and $g_2$ from~\Ref{Theta} in
\Ref{LYM} correspond to deformations of the ungauged theory of two
different types: The constant $g_1$ arises as gauge coupling constant
of the gauge group ${\rm G}_0$ while $g_2$ appears as proportionality
factor of the Chern-Simons term which may be viewed as another
deformation of the ungauged YM theory. In addition, both constants
appear in the $T$-tensor and thereby in the scalar potential~$W$ which
is a quadratic polynomial in $g_1$, $g_2$. Let us however stress once
more that in general $g_1$ and $g_2$ are not free parameters
in~\Ref{Theta} but related by some consistency relation implied by
supersymmetry. In the degenerate case $g_1=0$, the dualized theory
\Ref{LYM} appears with an abelian gauge group $U(1)^\nu$ which does
not act on the scalar fields, and is deformed only by the presence of
the Chern-Simons term and a scalar potential. This has been worked out
in \cite{BeHaSa02} in the context of the $N=2$ theories describing the
Calabi-Yau fourfold compactifications of M-theory with
flux~\cite{HaaLou01}.

\section{Coupling massive vector fields}
\label{MVF}

The above elimination procedure can be extended in a straightforward
fashion to include couplings to massive vector fields in the framework
of pure CS gaugings~\Ref{LCS}. As we will now explain these massive
vector fields correspond to additional nilpotent directions in the
CS gauge group.

To this aim we consider an extension of the CS gauge group ${\rm
G}_0 \ltimes {\rm T}_\nu$ by a set $\hat{\rm T}_p$ of $p$ nilpotent
generators transforming in some representation of ${\rm G}_0$ and
closing into ${\rm T}_\nu$. Accordingly, the Lie algebra relations
\Ref{algebra} are extended by
\ba
\left[ \CJ^m ,\hat\GT^\Ga \right] ~=~ t^{m\Ga}{}_\Gb\,
    \hat\GT^\Gb \;,\qquad
\left[ \hat\GT^\Ga ,\hat\GT^\Gb \right] ~=~
t^{\Ga\Gb}{}_m\,\GT^\Um\;,\qquad
\left[ \GT^\Um ,\hat\GT^\Ga \right] ~=~ 0 \;,
\la{algebra_ext}
\ea
while the embedding tensor $\GTh_{\cM\cN}$ has the additional
components
\ba
g\,\GTh_{\Ga\Gb} &=& g_1\, \kappa_{\Ga\Gb} \;,
\la{Theta_ext}
\ea
with the structure constants in \Ref{algebra_ext} and the symmetric
tensor $\kappa_{\Ga\Gb}$ being related by $\eta_{mn}t^{n\Ga}{}_\Gb =
\kappa_{\Gb\Gg} t^{\Ga\Gg}{}_m$ in order to have $\GTh_{\cM\cN}$
invariant under the gauge group. We denote the group corresponding to
\Ref{algebra}, \Ref{algebra_ext} by ${\rm G}_0 \ltimes (\hat{\rm T}_p,
{\rm T}_\nu)$. In addition to the vector fields~\Ref{gtvec} there are
now also vector fields $B_\mu^\Ga$ corresponding to the nilpotent
generators $\hat\GT^\Ga$. Similar to \Ref{Stilde}, we may also split
off the scalars associated with the generators $\hat\GT^\Ga$ from
the scalar field matrix $\CS$
\ba
\CS (\phi, \tilde\phi) &\equiv& e^{\phi_m\,\GT^{\Um}}\,
e^{\phi_\Ga\,\hat\GT^{\Ga}}\, \TCS (\tilde\phi) \;.
\la{STT}
\ea
Explicitly, the individual parts of \Ref{STT} transform under gauge
transformations as
\ba
\TCS &=& - g_1 \eta_{mn} \Lambda^{\Um}\, \CJ^n \TCS \;,
\qquad \delta \phi_\Ga ~=~ - g_1 \kappa_{\Ga\Gb} \Lambda^\Gb + g_1
t_{m\Ga}{}^\Gb \phi_\Gb \,\Lambda^\Um
\;,
\la{gauge_ext}\\
\delta \phi_m &=& - \eta_{mn} g_1 \Lambda^n
-\ft12 g_1 t_{m\Ga}{}^\Gb \phi_\Gb \Lambda^\Ga-
\left( g_2 \eta_{mn}  + g_1 f^l{}_{mn}\phi_l
-\ft12 g_1 t^{\Ga\Gg}{}_m t^\Gb{}_{\Gg n}\,\phi_\Ga\phi_\Gb
\right) \Lambda^\Un \;,
\nn
\ea
and correspondingly this defines their covariant derivatives. The
elimination procedure described in the last section may now
straightforwardly be generalized to this setting. We refrain from
giving details of the computation and just note that \Ref{PQtild}
generalizes to
\ba
\CP^A_\mu &=&
\TCP^{A}_\mu + \TCV^\Ga{}_A \, D_\mu\phi_\Ga
+ \TCV^\Um{}_A\,(D_\mu\phi_m -
\ft12 t^{\Ga\Gb}{}_m \phi_\Ga D_\mu \phi_\Gb)
\;,
\la{Ptild_ext}
\ea
with $\TCV$, $\TCP$ defined as in \Ref{VT}. The first order duality
equations~\Ref{duality} then allow to express $D_\mu\phi_m$ and the
field strength $\CC^m_{\mu\nu}$ in terms of the remaining fields,
thereby eliminating them from the theory. As above, integrability of
these equations implies a second order field equation for the vector
fields $A_\mu^m$ that generalizes~\Ref{YM}. In addition, we remain
with a first order equation for the vector fields $B_\mu^\Ga$
associated with the nilpotent generators $\hat{\GT}^\Ga$
\ba
\TCB_{\mu\nu}^\Ga &\equiv&
D_\mu B^\Ga_\nu - D_\nu B^\Ga_\mu -
t^{\Ga\Gb}{}_m\,\phi_\Gb\,A^m_{\mu\nu}
\non[1ex]
&=&
\TCV^\Ga{}_A \TCV^\Um{}_A\,M_{mn}\,A^n_{\mu\nu} +
e\Ge_{\mu\nu\rho} \TCV^\Ga{}_A \,G_{AB}\,
(\TCP^{B\rho} + \TCV^\Gb{}_B\,D^\rho\phi_\Gb ) \;,
\la{dualB}
\ea
with $G_{AB}$ from \Ref{LYMa}, and where left and right hand side are
separately invariant under gauge transformations with parameters
$\Lambda^\Ga$, $\Lambda^m$. Finally, the entire set of field equations
may be derived from the Lagrangian
\ba
e^{-1}\TCL &=& -\ft1{4} R
+ \ft1{4} G_{AB}\,
(\TCP^A_\mu + \TCV^\Ga{}_A\,D_\mu\phi_\Ga )
(\TCP^{B\mu} + \TCV^\Gb{}_B\,D^\mu\phi_\Gb )
-\ft18 \,M_{mn}\,\CA^{m}_{\mu\nu}\CA^{n\mu\nu}
\non[1ex]
&&{}
+\ft14 e^{-1} \,
\Geps^{\mu\nu\rho}\,M_{mn}\TCV^\Un{}_A\,\CA^{m}_{\mu\nu} \,
(\TCP^A_\rho + \TCV^\Ga{}_A\,D_\rho\phi_\Ga )
-\ft18 e^{-1} \,
\Geps^{\mu\nu\rho}\,\TCB^\Ga_{\mu\nu}\,D_\rho\phi_\Ga
\non[1ex]
&&{}+ e^{-1}g_2 \TCL_{\rm CS}(A) - W
\;,
\label{LYM_ext}
\ea
which generalizes \Ref{LYM} by including a coupling to the additional
vector fields $B^\Ga_\mu$ which arise with the first order field
equation \Ref{dualB}. The Lagrangian~\Ref{LYM_ext} has the additional
gauge symmetry $\Lambda^\Ga$ exclusively acting on $B^\Ga_\mu$ and
shifting $\phi_\Ga$. In particular, this symmetry may be gauge fixed
by imposing $\phi_\Ga=0$ which leads to the Lagrangian
\ba
e^{-1}\TCL &=& -\ft1{4} R
+ \ft14 G_{AB}\,
\TCP^A_\mu
\TCP^{B\mu}
-\ft18 \,M_{mn}\,\CA^{m}_{\mu\nu}\CA^{n\mu\nu}
+\ft14 e^{-1}
\Geps^{\mu\nu\rho}\,M_{mn}\TCV^\Un{}_A\,\CA^{m}_{\mu\nu} \,
\TCP^A_\rho
\non[1ex]
&&{}
+ \ft1{4} \,g_1^2\,\TCV_{\Ga A}\,G_{AB}\,\TCV_{\Gb B}\,
B_\mu^\Ga B^{\Gb\mu}
+\ft14g_1\, e^{-1}
\Geps^{\mu\nu\rho}\,M_{mn}\TCV^\Un{}_A\TCV_{\Ga A}\,
\,B^\Ga_\rho\,\CA^{m}_{\mu\nu}
\label{LYM_exta}
\\[1ex]
&&{}
+ \ft1{2} g_1 \, G_{AB}\TCV_{\Ga B}\,
\TCP^A_\mu  \, B^{\Ga\mu}
-\ft18 g_1\,e^{-1}
\Geps^{\mu\nu\rho}\,B^\Ga_\rho\,\kappa_{\Ga\Gb}\,\CB^\Gb_{\mu\nu}
+ e^{-1} g_2 \TCL_{\rm CS}(A) - W
\;.
\nn
\ea
This is a theory describing $(d-\nu-p)$ scalar fields combined in the
matrix $\TCS$, together with $p$ massive CS vector fields $B^\Ga_\mu$,
and $\nu$ YM vector fields $A^m_\mu$ gauging the group $G_0$ with the
gauge symmetry acting as
\ba
\delta A^m_{\mu} &=&
\dd_\mu \Lambda^{\Um} +g_1\,f^m{}_{kl}\,A^k_{\mu}\,\Lambda^\Ul \;,
\non[.5ex]
\delta \TCS &=& - g_1 \eta_{mn} \Lambda^{\Um}\, \CJ^n \TCS \;,
\qquad
\delta B^\Ga_{\mu} ~=~
-g_1\,t^\Ga{}_{m\Gb}\,B^\Gb_{\mu}\,\Lambda^\Um \;.
\ea
Lagrangians of the type \Ref{LYM_exta} typically arise in
dimensional reduction on nontrivial internal
manifolds~\cite{CvLuPo00,LuPoSe02}, see the following section.
Note that in contrast to \Ref{LCS}, \Ref{LYM_ext}, this Lagrangian
no longer has a smooth limit $g_1\rightarrow0$ because some
propagating degrees of freedom decouple with the $B^\Ga_\mu$. It
is worthwhile to emphasize the simplicity of the original
Lagrangian \Ref{LCS} in comparison with \Ref{LYM_exta}, which is
due to the fact that all the different scalar tensors which
describe the couplings of the various fields in~\Ref{LYM_exta} are
encoded in the original ${\rm G}/{\rm H}$ coset structure.
Moreover, the gauge deformation of \Ref{LCS} is uniformly
described in terms of the embedding tensor $\Theta$ which
transforms covariantly under the maximal global symmetry ${\rm G}$
underlying the CS description of the gauged theory.

\section{Examples}
\la{EXS}

Having shown that the CS gauged supergravities \Ref{LCS} contain the
YM type theories \Ref{LYM} and \Ref{LYM_exta} as special cases, we
now have the means to construct any three-dimensional supergravity
given the number of supersymmetries, gauge group and field content.
To do so, one first identifies that version of the underlying ungauged
supergravity for which all propagating bosonic degrees of freedom appear
as scalar fields and are uniformly described by a maximal coset space
${\rm G}/{\rm H}$. By contrast, the YM type theory is based on a
description where only part of the bosonic degrees of freedom correspond
to scalar fields, such that \Ref{matching} holds.

Together with the precise representation content under the gauge group
and for sufficiently large number $N$ of supersymmetries, this is
already sufficient to identify the corresponding theory in the list
of~\cite{dWToNi93}. Next, the gauge group must be chosen as a
subgroup of ${\rm G}$ such as to reproduce the correct representation
content while its non-semisimple part determines the nature of the
vector couplings as explained in sections~\ref{CSYM},~\ref{MVF}. In
addition, the embedding tensor $\GTh_{\cM\cN}$ of this group is
constrained by the group-theoretical consistency condition
of~\cite{NicSam00,NicSam01b,dWHeSa03} where the complete form of the
CS gauged theory is then found. Finally, one may apply the
constructions presented in this paper to cast the vector couplings
into the desired form \Ref{LCS},~\Ref{LYM}, or~\Ref{LYM_exta}.

\bigskip

We conclude with some examples that reproduce the
mass spectra and symmetries of known AdS$_3$ compactifications.
Most of these theories have not been constructed before. Recall
that in dimensional reduction one typically encounters YM gauged
theories, i.e.\ the Lagrangian obtained directly by compactification
will take the form \Ref{LYM}, \Ref{LYM_exta} rather than the equivalent
simpler form \Ref{LCS}. As pointed out in \cite{NicSam00} the gauged
CS type theories with semisimple gauge groups have no obvious higher
dimensional ancestor. It is therefore remarkable that the CS type models
which can be linked to higher dimensional supergravities, all have
non-semisimple gauge groups. Could there be a higher-dimensional theory
that gives rise to these models in a singular limit akin to the
boost limit producing non-semisimple gaugings from semisimple ones?
We should also stress that at this stage we restrict attention
to the (unique) lower-dimensional theories
with the correct field content, and are not concerned with the
consistency of the truncations from the higher-dimensional
point of view.

\begin{itemize}

\item[$-$]
One of the main examples of the AdS/CFT correspondence is the duality
between type IIB string theory on AdS$_3 \times S^3\times M^4$ and
certain two-dimensional conformal field theories~\cite{Mald97}. The
spectrum of $N=(2,0)$ supergravity on AdS$_3 \times S^3$ has been
computed in~\cite{DKSS98,Lars98,dBoe99}, in three dimensions this is a
half-maximal, i.e.\ $N=8$ theory. In~\cite{NicSam01b} we have shown
that the lowest multiplets of this spectrum together with the expected
global and local symmetries are reproduced by a three-dimensional
theory~\Ref{LCS} with coset space $\SO{8,n}/(\SO{8}\cro\SO{n})$, and
CS gauge group $\SO{4}$, where $n$ denotes the number of tensor
multiplets in six dimensions. An outstanding question has been the
coupling of this theory to the YM vector multiplet containing
additional 26 scalars. In view of the above results, one may now
verify that this larger theory may be described by a coset space
$\SO{8,4\pls n}/(\SO{8}\cro\SO{4\pls n})$ and CS gauge group
$\SO{4}\ltimes {\rm T}_6$. This $\SO{4}$ is embedded as a certain
diagonal of two factors in the $\SO{8}$ and the $\SO{4\pls n}$,
respectively, such that the scalar spectrum decomposes as $\big(8,
(n\pls4)\big)\rightarrow n\cdot4 + 4n\cdot1 + 1+9+4\cdot4 + 3_+ +
3_-$. Eliminating the six abelian translations as described in
section~\ref{CSYM} leads to a YM $\SO{4}$ gauged theory~\Ref{LYM}
coupling 26+8$n$ scalar fields.  Details of this construction will be
presented in~\cite{NicSam03b}. Surprisingly, using the results of
section~\ref{MVF} and a particular coset space, it is even possible to
describe the coupling of multiplets from arbitrary~(!)  levels of the
massive spin-1 KK towers.

\bigskip

\item[$-$]
The near horizon limit of the so-called ``double D1-D5 system''
describes an AdS$_3 \times S^3 \times S^3$ geometry. The supergravity
spectrum on this background has been computed in~\cite{dBPaSk99}. The
three-dimensional theory describing the lowest mass multiplets again
is organized by a coset space $\SO{8,n}/(\SO{8}\cro\SO{n})$, now with
gauge group $\SO{4}\times\SO{4}$~\cite{NicSam01b}. Similar to the
construction given above, one may further couple the two YM multiplets
by a proper enlargement of the coset space to
$\SO{8,8+n}/(\SO{8}\cro\SO{8+n})$ and embedding a gauge group
$\SO{4}_{\rm diag}\times\SO{4}_{\rm diag}$ together with the
corresponding nilpotent directions.

\item[$-$]
Recently, the reduction of of six-dimensional $N=(1,0)$ supergravity
on AdS$_3 \times \SU{2}$, has been performed in~\cite{LuPoSe02}. The
field content of the three-dimensional $N=4$ theory comprises three YM
gauge fields together with three massive vector fields and six scalars
parametrizing the coset space ${\rm GL}(3)/\SO{3}$. This spectrum
suggests that the CS version \Ref{LCS} of this theory is governed by
the larger coset space $\SO{4,3}/(\SO{4}\times\SO{3})$ with gauge
group $\SO{3}_{\rm diag}\ltimes (\hat{\rm T}_{3}, {\rm T}_{3})$. The
group $\SO{4,3}$ indeed has a unique subgroup of this type, whose
algebra generators satisfy relations of the type \Ref{algebra},
\Ref{algebra_ext}. Its semi-simple part is embedded as the diagonal of
the three $\SO{3}$ factors in the compact $\SO{4}\times\SO{3}$, such
that the 12 scalars decompose as $12\rightarrow 1+3+3+5$. Eliminating
the three abelian translations and gauge-fixing the other three
nilpotent directions as described in section~\ref{MVF} leads to a
theory~\Ref{LYM_exta} with the desired spectrum. Indeed, the couplings
appearing in~\Ref{LYM_exta} for this case are of the form
found in~\cite{LuPoSe02}.

\item[$-$]
The reduction of six-dimensional $N=(1,0)$ supergravity on
AdS$_3 \times S^3$ leads to a three-dimensional $N=4$ theory, whose
spectrum has been computed in~\cite{DKSS98,dBoe99}. The resulting
theory will be described by a coset space
$\SO{4,4}/(\SO{4}\cro\SO{4})$ and CS gauge group $\SO{4}_{\rm
diag}\ltimes {\rm T}_{6}$ embedded such that the scalar spectrum
correctly decomposes as $16\rightarrow 1+ 3_+ + 3_- + 9$, of which the
$3_+ + 3_-$ transforming in the adjoint representation of $\SO{4}_{\rm
diag}$ are eliminated according to section~\ref{CSYM}.

\item[$-$]
The compactification of simple five-dimensional supergravity on $S^2$
whose KK spectrum has been analyzed in \cite{dBoe99,FuKeMi98,Suga99}
should be related to proper gaugings of the $N\equ4$ theory with coset
space~$\EG22/\SO{4}$. The CS gauge group is a $\SO{3}_{\rm
diag}\ltimes {\rm T}_{3}$ under whose semisimple part the scalar
spectrum decomposes as $8\rightarrow 3+5$.

\item[$-$]
The most interesting example is the (warped) compactification of
ten-dimensional supergravity on~$S^7$. For the type I theory, the
reduction has been performed explicitly in~\cite{CvLuPo00}. We may
recover that theory by starting from an $N=8$ theory \Ref{LCS} with
coset space $\SO{8,8}/(\SO{8}\times\SO{8})$, and CS gauge group
$\SO{8}_{\rm diag}\ltimes {\rm T}_{28}$ upon eliminating the $28$
translations, leading to an $\SO{8}$ gauged YM theory~\Ref{LYM} with
$36$ scalar fields. Using the above results, it is then
straightforward to extend this construction to the maximally
supersymmetric theories which are supposed to describe the
compactifications of type IIA/IIB theory on $S^7$, and whose spectra have
been given in~\cite{MorSam02}. These two  different compactifications
correspond to two inequivalent embeddings of the gauge group
$\SO{8}_{\rm diag}\ltimes {\rm T}_{28}$ into $\EE88$.
Details will appear in \cite{FiNiSa03}.

\end{itemize}

\subsection*{Acknowledgements}

It is a pleasure to thank M.~Berg, B.~de Wit, T.~Fischbacher, and
M.~Haack for discussions. This work is partly supported by EU contract
HPRN-CT-2000-00122.


\renewcommand{\baselinestretch}{1}

{\small

\providecommand{\href}[2]{#2}
\begingroup\raggedright\endgroup

}

\end{document}